\documentclass[aps,prl,twocolumn,superscriptaddress,showpacs]{revtex4-1}
\usepackage{graphicx}
\usepackage{siunitx}
\usepackage{amssymb}
\begin{document}

\title{Bistable Transport Properties of a Quasi-One-Dimensional Wigner Solid on Liquid Helium under Continuous Driving}


\author{David G. Rees}
\email[]{drees@nctu.edu.tw}
\affiliation{NCTU-RIKEN Joint Research Laboratory, Institute of Physics, National Chiao Tung University, Hsinchu 300, Taiwan}
\affiliation{RIKEN CEMS, Wako 351-0198, Japan}
\affiliation{KFU-RIKEN Joint Research Laboratory, Institute of Physics, Kazan Federal University, Kazan, 420008 Russia}

\author{Sheng-Shiuan Yeh}
\affiliation{NCTU-RIKEN Joint Research Laboratory, Institute of Physics, National Chiao Tung University, Hsinchu 300, Taiwan}

\author{Ban-Chen Lee}
\affiliation{NCTU-RIKEN Joint Research Laboratory, Institute of Physics, National Chiao Tung University, Hsinchu 300, Taiwan}

\author{Kimitoshi Kono}
\affiliation{NCTU-RIKEN Joint Research Laboratory, Institute of Physics, National Chiao Tung University, Hsinchu 300, Taiwan}
\affiliation{RIKEN CEMS, Wako 351-0198, Japan}
\affiliation{KFU-RIKEN Joint Research Laboratory, Institute of Physics, Kazan Federal University, Kazan, 420008 Russia}

\author{Juhn-Jong Lin}
\affiliation{NCTU-RIKEN Joint Research Laboratory, Institute of Physics, National Chiao Tung University, Hsinchu 300, Taiwan}
\affiliation{RIKEN CEMS, Wako 351-0198, Japan}
\affiliation{Department of Electrophysics, National Chiao Tung University, Hsinchu 300, Taiwan}

\date{\today}
\begin{abstract}
We investigate low-frequency fluctuations in the transport characteristics of a quasi-1D Wigner solid (WS) moving along a liquid helium substrate in response to a sinusoidal driving voltage. The fluctuations occur between distinct transport modes in which the decoupling of the WS from ripplonic polarons (or `dimple lattice', DL) formed on the helium surface does, or does not, occur during each ac cycle. We demonstrate that a Gaussian-like distribution in the decoupling threshold force gives rise to this bistability, as the low-frequency switching occurs when the probability of decoupling during each ac cycle is small but finite. We attribute the distribution in the decoupling threshold force to the range of structural configurations allowed for the quasi-1D electron lattice, which influences the strength of the WS-DL coupling. Hence, the switching rate between the ac transport modes is extremely sensitive to the microscopic properties of the electron solid. 
\end{abstract}

\maketitle

\section{I. Introduction}

Bistable behaviour has been studied extensively both experimentally and theoretically in a wide variety of physical systems including electromechanical resonators \cite{cantilever,DykmanBistability}, Josephson junctions  \cite{Fulton} and trapped electrons and atoms \cite{Gabrielse,PhysRevLett.64.408}. Bistability can also be induced or triggered in dynamical systems by parametric driving \cite{Imran,Siddiqi,Lapidus,DykmanPRE}. Here we demonstrate a novel bistable process, observed in the transport of electron crystals trapped on the surface of liquid helium and driven back and forth through a narrow channel. Transport modes in which the electrons are coupled to or decoupled from helium surface excitations provide two `states' between which a stochastic switching can occur. We demonstrate that this effect arises due to a distribution in threshold force at which the electron crystal decouples from the helium surface. As well as revealing new information regarding the interaction between the electron crystal and its liquid substrate, our experiment provides a clear example of how the application of a continuous driving force can invoke and elucidate metastability in dynamical systems.     

At sufficiently low temperatures, electrons trapped on the surface of liquid helium \cite{Andrei,MonarkhaKono} self-organise to form a triangular lattice, the Wigner solid (WS) \cite{Wigner,GrimesAdamsWignerCrystal}. Unlike the WS states found in degenerate 2D electron systems in the quantum regime \cite{Csathy}, the WS on helium is essentially classical in nature due to the relatively low electron densities ($10^9\lesssim n_s \lesssim 10^{13}$ m$^{-2}$) that can be supported by the helium substrate. In addition, the Coulomb interaction between the electrons is not screened by the helium, which has a permittivity close to that of vacuum. This makes surface electrons an ideal model system for the study of strongly correlated electron behaviours in 2D \cite{GrimesAdamsMobility} and, in recent experiments using microchannel devices, transport at the nanoscale in the limit of strong electron-electron interaction \cite{PointContact,Bradbury}.

The interaction between the electron lattice and the liquid-vacuum interface gives rise to several intriguing phenomena. The pressure exerted on the helium surface by each electron results in a shallow deformation of the liquid, beneath each electron, which is known as a `dimple' \cite{MonarkhaShikinDimple,Fisher1979}. Equivalently, and in analogy with polaron states formed by electrons coupled to virtual phonons in crystal lattices, the static WS can be described as being `dressed' by quantised capillary waves (ripplons) \cite{Jackson1981,saitoh1986}. Because the combined mass of the electron and its dimple is much greater than the bare electron mass, the conductance of the electron system usually drops dramatically when the system enters the WS phase \cite{GIANNETTA1991199}. However, the WS can be decoupled from the surface deformation by the application of a strong electric field parallel to the helium surface \cite{PhysRevLett.74.781,*Shirahama2}. After decoupling, the WS moves freely across the helium with high velocity. Analogous decoupling processes occur for polarons formed in solid-state systems \cite{alexandrov2010advances}.

\begin{figure*} [t]
\includegraphics[angle=0,width=0.6\textwidth]{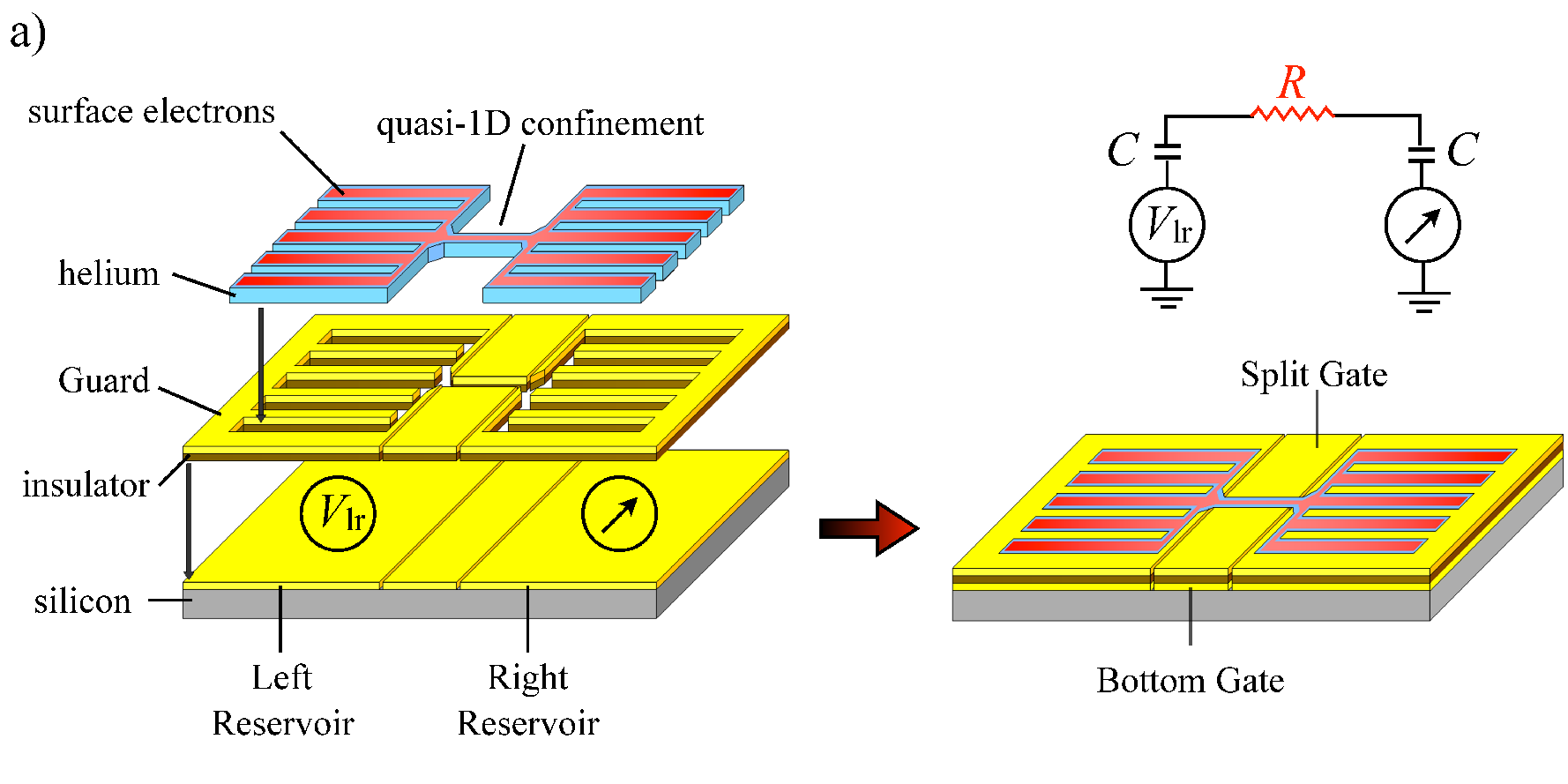}
\includegraphics[angle=0,width=0.3\textwidth]{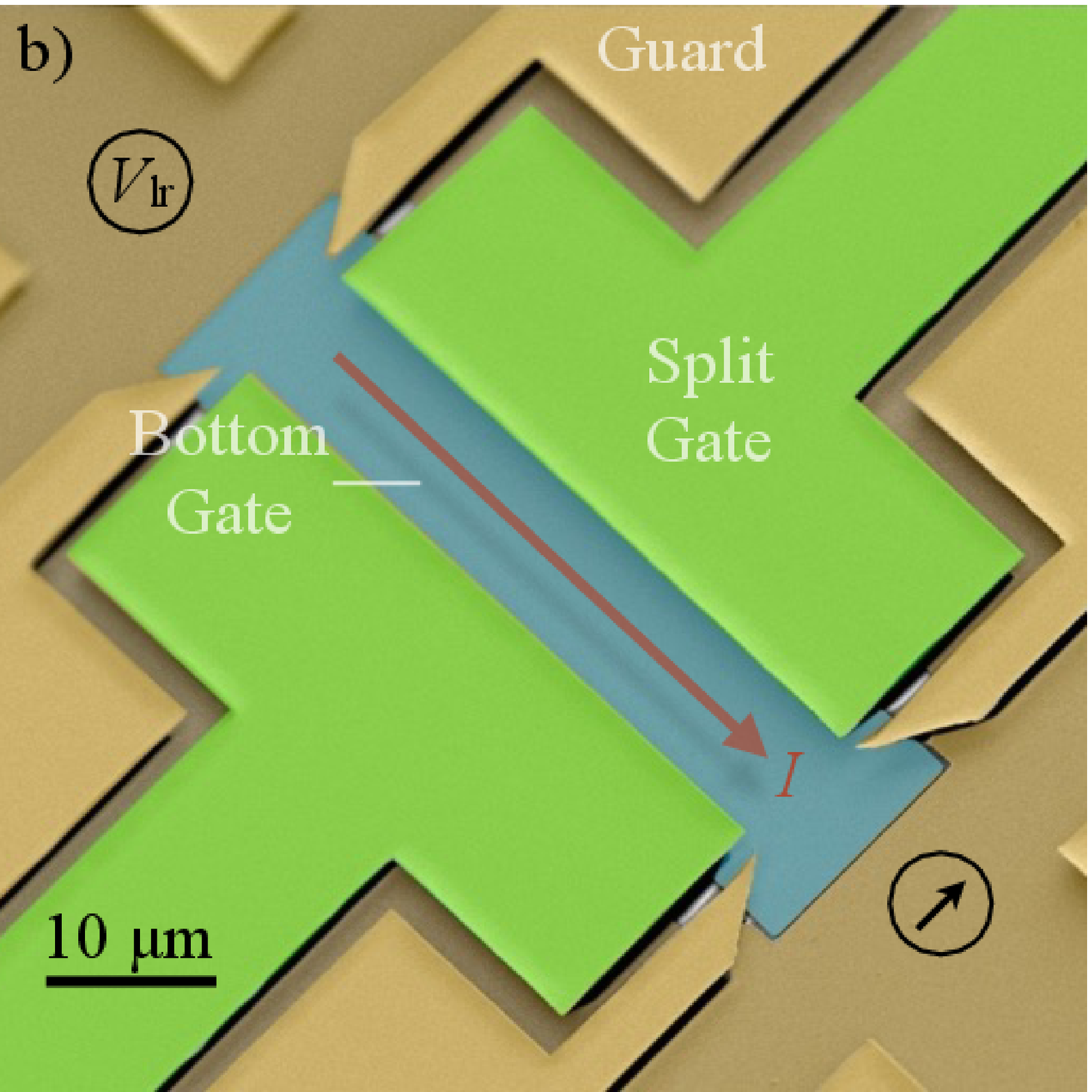}
\caption{(a) Schematic picture of the microchannel device. The left and right reservoir and bottom gate electrodes in the lower layer are separated from the guard and split gate electrodes in the upper layer by hard-baked photoresist. Liquid helium fills the channel structures by capillary action. Electrons are then deposited on the helium surface. The device can be represented by the simplified RC circuit diagram. (b) False-colour scanning electron micrograph of the device. The surface electron current $I$ flows through the central microchannel in response to the driving voltage $V_{lr}$ and induces a displacement current in the right reservoir electrode.\label{Fig:1}}
\end{figure*} 

In recent time-resolved transport measurements, a repeated decoupling of the WS from the dimple lattice (DL) was observed for a WS moving through a helium-filled microchannel a few microns in width, in response to a smoothly ramped driving voltage \cite{Stick-slip}. The consequent `stick-slip' motion of the electron system resulted in spontaneous current oscillations, the frequency of which was dependent on the strength of coupling between the WS and the helium substrate and on the applied driving force. However, many details regarding the decoupling process remain poorly understood. In this report we present new transport measurements that uncover important information regarding the WS-DL decoupling. We demonstrate that, during a driving voltage ramp, the threshold electric field at which the decoupling can occur follows a Gaussian-like distribution. We then investigate the influence of this threshold distribution on the WS response to a sinusoidal driving voltage. In this case, two transport modes are possible; in the first mode the WS-DL decoupling occurs once during each half of the ac cycle, whilst in the second mode the WS and DL always remain coupled. Because, under continuous driving, the response during one cycle influences the behaviour during the next, the system can become `locked' in one mode or the other. However, the distribution in the decoupling threshold force allows the system to occasionally `switch' between the modes, resulting in telegraph-like `noise' in the current amplitude. By tuning the system into a regime in which these fluctuations are rare, the average time between switching events becomes much longer than the ac driving period. The low-frequency noise observed in our experiment can reveal information regarding the microscopic details of the quasi-1D electron crystal. 

\section{II. Experimental}

The sample, which is shown in Fig. 1, and the measurement set up used in the experiments have been described in detail elsewhere \cite{beysengulov2016structural,PhaseDiagram}. The sample is fabricated using UV lithography on a section of silicon wafer. Two Au layers, separated by a 2.2 $\mu$m-thick layer of insulating hard-baked photoresist, are patterned to form the various electrodes used to confine and manipulate the electron system. In the upper layer, the guard electrode forms the perimeter of the sample and defines two arrays of microchannels, in which each microchannel is 20 $\mu$m wide, that are used to store a large number ($\sim10^6$) of electrons. These regions are denoted the left and right reservoirs. A small central microchannel connects the two reservoirs; its geometry is defined by the split-gate electrode. The central microchannel is 7.5 $\mu$m wide and 100 $\mu$m long. In the lower metal layer, the bottom gate electrode covers the area at the bottom of the central microchannel, and the left and right reservoir electrodes cover the area at the bottom of the left and right reservoirs, respectively. The microchannels are filled by the capillary action of superfluid helium when the liquid surface is close to the sample. Surface electrons are then generated by thermionic emission from a tungsten filament situated a few mm above the sample.

To trap electrons in the reservoirs and to control their transport through the central microchannel, dc voltages $V_{gu}=-0.2$ V, $V_{sg}$ and $V_{bg}$ were applied to the guard, split gate and bottom gate electrodes, respectively. Transport measurements were made by applying ramped or sinusoidal driving voltages to the left reservoir electrode using an arbitrary waveform generator; the time-dependent voltage on the left reservoir is denoted $V_{lr}$. The displacement current induced in the right reservoir electrode (which is held at dc ground) was then measured using a current preamplifier and either a lock-in amplifier or a digital storage oscilloscope. For the time-resolved measurements of current fluctuations, the demodulated signal from the lock-in amplifier was monitored using a spectrum analyser. All measurements were made at temperature $T=0.6$ K, unless otherwise stated.

\section{III. Response to a Voltage Ramp}

We start by analysing the response of the quasi-1D WS to a smooth voltage ramp. In Fig. 2(a) we show the surface electron current $I$ that flows in response to a change in $V_{lr}$ from 0 to +50 mV in 80 $\mu$s. Here $V_{sg}=-1$ V and $V_{bg}=1.14$ V. According to finite element modelling (FEM) \cite{beysengulov2016structural,hecht2012new}, under these bias conditions the effective width of the quasi-1D electron system is $w_e= 4.0$ $\mu$m, $n_s=2.3\times10^{13}$ m$^{-2}$, and the electron lattice comprises approximately 18 electron rows. The current was measured using a preamplifier (Femto DHPCA-100) with wide bandwidth (3.5 MHz). To resolve the signal above the experimental noise, the signal was averaged some 8000 times. 
\begin{figure} 
\includegraphics[angle=0,width=0.45\textwidth]{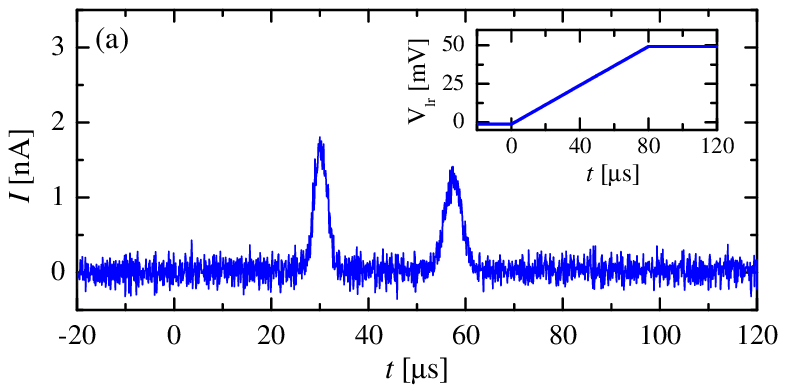}
\includegraphics[angle=0,width=0.45\textwidth]{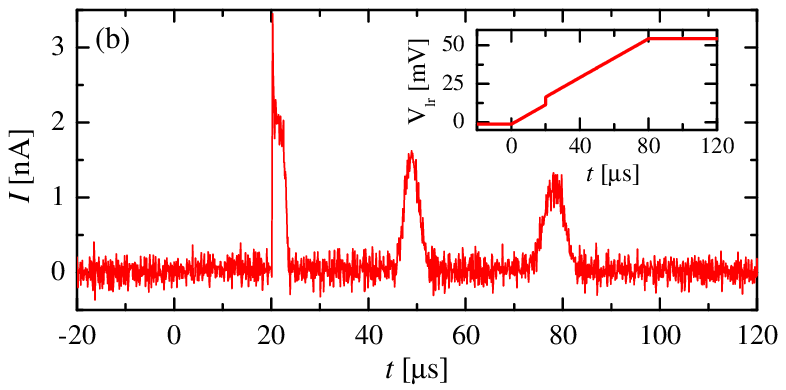}
\includegraphics[angle=0,width=0.45\textwidth]{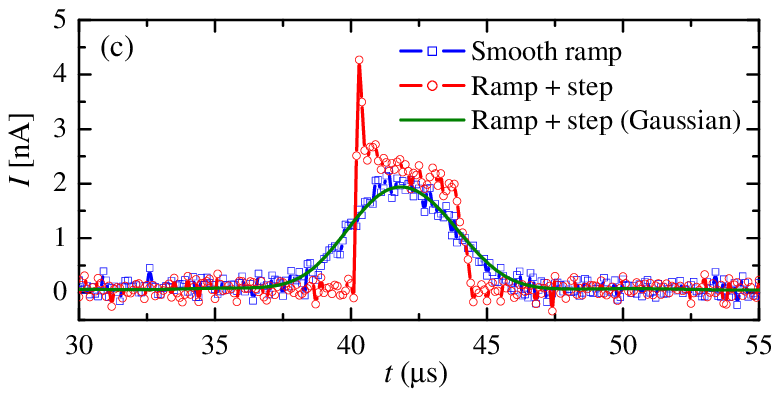}
\caption{(a) $I$ against $t$ for a smooth $V_{lr}$ ramp from 0 to +50 mV in 80 $\mu$s, as shown in the inset. Here $V_{sg}=-1$ V and $V_{bg}=1.14$ V. (b) $I$ against $t$ for the same conditions  but with the addition of a step of +5 mV to the $V_{lr}$ ramp at $t=20$ $\mu$s, as shown in the inset. (c) $I$ against $t$ for a smooth $V_{lr}$ ramp recorded for $V_{bg}=1.3$ V (squares). For this condition the current peak appears after approximately 50 $\mu$s. However, in the plot the data is shifted by $-9.2$ $\mu$s in order to allow comparison with the current recorded when a +5 mV voltage step is added to the $V_{lr}$ ramp at $t=40$ $\mu$s (circles). The solid line shows the result of simulating the effect of a Gaussian distribution in the WS-DL decoupling threshold, as described in the text.  \label{Fig:2}}
\end{figure} 

Two current peaks are recorded as $V_{lr}$ increases, due to the repeated decoupling of the WS from the DL. The period separating the decoupling events, and the shape of the resulting current peaks, is consistent with previous measurements \cite{Stick-slip}.
The WS-DL decoupling occurs due to the building of the driving electric field $E_x$ that acts on the electron system when $V_{lr}$ is ramped. At the beginning of the ramping phase, the WS moves with the DL. As the velocity of the system approaches the phase velocity $v_r$ of ripplons whose wavevector matches the first reciprocal lattice vector of the electron lattice, the moving WS excites ripplons which interfere constructively with the DL. As a result, the dimples become deeper and the drag force acting on the WS increases. The WS therefore becomes dynamically pinned to the DL, at velocity $v_r$. This process is known as Bragg-Cherenkov (BC) scattering \cite{DykmanRubo}. Because the resonant ripplon wavevector is well-defined (by the periodicity of the electron lattice) this velocity can be readily determined from the ripplon dispersion relation, if the electron density is known. The ripplon dispersion relation is $\omega^2=(\alpha/\rho)k^3$ where $\omega$ is the ripplon angular frequency, $\alpha$ is the helium surface tension, $\rho$ is the liquid density, and $k$ is the ripplon wavevector that, under the BC resonance, is equal to the first reciprocal lattice vector of the electron solid $G_1$. For the conditions in Fig. 2 the velocity at which the BC resonance occurs is $v_r=\omega/G_1=8.2$ m s$^{-1}$, which results in a current $I_{BC}=en_sw_ev_r=100$ pA (close to the noise floor of the measurement). However, in the limit of zero electrical resistance the WS would travel at a greater velocity determined by the $V_{lr}$ ramp rate, the capacitance of the electron sheet to the reservoirs $C=1.0$ pF, $w_e$, and $n_s$; we estimate this velocity to be 23.3 m s$^{-1}$. It is the reduced velocity of the coupled WS-DL system that leads to the increase in $E_x$ as $V_{lr}$ is ramped. As $E_x$ grows, a threshold value of $E_x$ is eventually reached at which the WS decouples from the DL, and the electrons move across the helium surface at a velocity much higher than $v_r$. This results in a burst of current that largely restores the electrostatic equilibrium, reducing the driving field. The WS can then recouple with the DL and the stick-slip cycle can start again. 

The distinct current oscillations due to the stick-slip WS motion have been described previously \cite{Stick-slip}. However, because signal averaging was necessary to resolve the current above the background noise, it was not certain that this measurement recorded the true dynamics of an individual decoupling event. To investigate this issue, the measurement shown in Fig. 2(a) was modified by adding a small voltage step of +5 mV (made in 0.1 $\mu$s) to the usual smooth voltage ramp, at $t=20$ $\mu$s (just before the first current peak appears in Fig. 2(a)). The response to the modified ramp is shown in Fig. 2(b). A current peak now appears at $t=20$ $\mu$s, with a quite different lineshape to those in Fig. 2(a). The current rises abruptly and then flows at an almost constant value, before decreasing rapidly again. Following this, two subsequent decoupling events occur, which generate current peaks similar to those observed in Fig. 2(a).

The addition of a sharp voltage step to the smooth ramp before, but close to, the point at which the WS decouples from the DL during the smooth voltage ramp ensures that the WS-DL decoupling is induced within the 0.1 $\mu$s step for each driving voltage cycle. In this case, although averaging is still used to resolve the current signal, the measured increase in the current at $t=20$ $\mu$s is extremely sharp; it follows that the current peaks observed when no voltage step is applied appear smooth due to an averaging of many decoupling events that each occur at slightly different times. To test this hypothesis, we assume that the distribution of the decoupling times is Gaussian in nature and described by a standard deviation $\sigma_t$. In Fig. 2(c) we take a typical example of the current signal synchronised by the small voltage step (in this case at $t=40$ $\mu$s) and simulate the effect of averaging many of these signals with a distribution in their decoupling times. For each data point, recorded at time $t_p$, we generate a Gaussian distribution centred about $t_p$, the sum of which has an amplitude equal to the value of that data point. We then sum these distributions. As shown in Fig. 2(c), the result is a smooth current peak that, for $\sigma_t=1.4$ $\mu$s, closely matches a typical example of a current peak recorded during the smooth voltage ramp (this data is shifted by -9.2 $\mu$s to allow comparison). We conclude that a Gaussian-like distribution in the times at which the WS decouples from the DL accounts for the smoothed appearance of the stick-slip current features. The current peaks recorded in response to the voltage step, which more accurately reflect the strongly nonlinear WS transport, will be examined in more detail elsewhere. 

The observation that the WS-DL decoupling occurs in a stochastic manner also explains the reduced amplitude of the second current peak in Fig. 2(a). Because the second decoupling events depend on the time at which the first occur, but also follow the same Gaussian-like distribution, the second peak appears more smoothed than the first. As expected, subsequent peaks become further broadened \cite{Stick-slip}.  

The stochastic nature of the WS-DL decoupling under a ramped potential indicates that the driving electric field at which the electron system decouples from the DL is not the same on each voltage cycle. For the measurement shown in Fig. 2(a), the rate of change of voltage across the electron system, before the decoupling occurs, is $dV_{lr}/dt-2I_{BC}/C\approx 445$ Vs$^{-1}$. The value of $\sigma_t$ can be used to characterise the distribution in driving voltage (and electric field) at which the decoupling occurs; $\sigma_t=1.4$ $\mu$s corresponds to a standard deviation in the voltage and electric field threshold distribution of $\sigma_V=0.62$ mV and $\sigma_{E_x}=6.2$ V/m, respectively. 

\section{IV. Response to a Sinusoidal Voltage}

\begin{figure} 
\includegraphics[angle=0,width=0.35\textwidth]{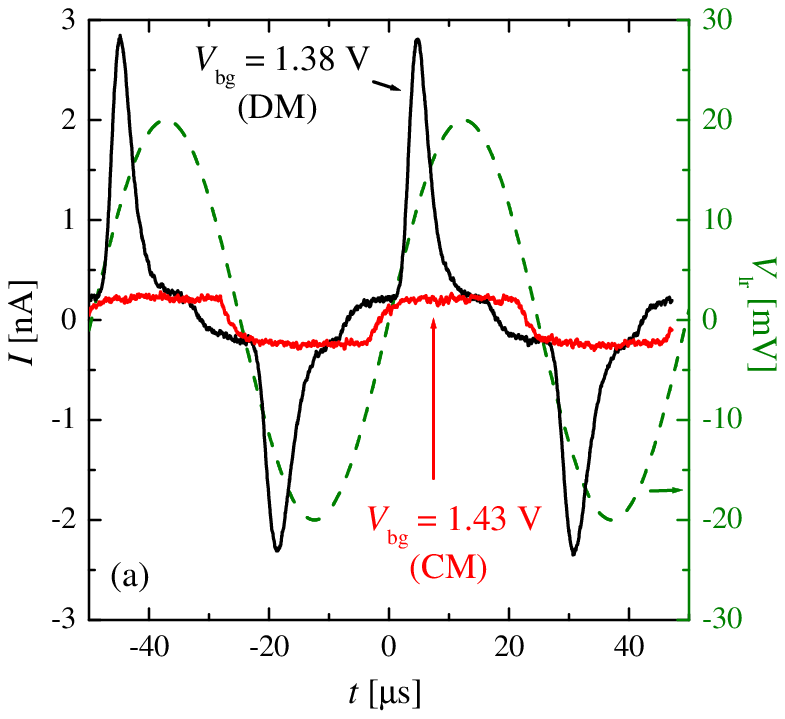}
\includegraphics[angle=0,width=0.35\textwidth]{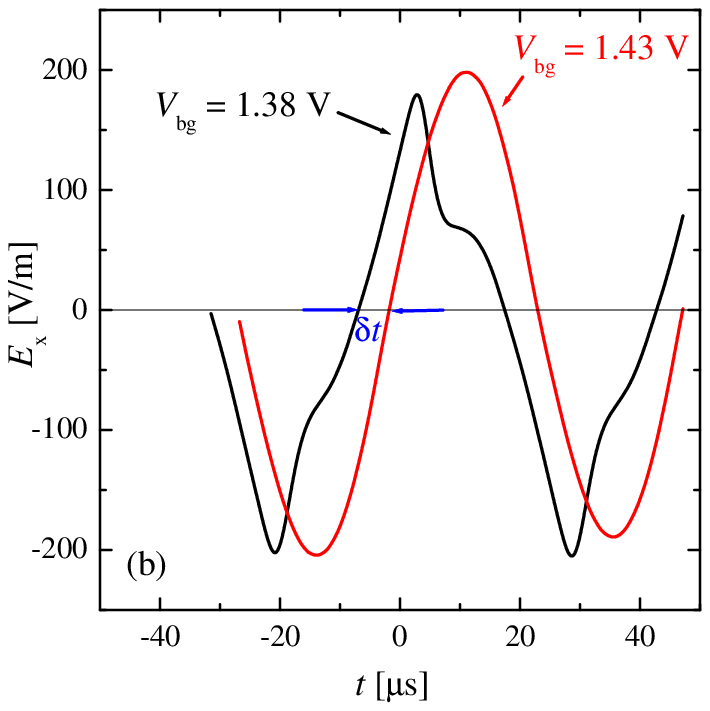}
\caption{(a) $I$ recorded for the DM and CM (solid lines) and the sinusoidal driving voltage $V_{lr}$ (dashed line) against $t$. Here $V_{sg}=-0.5$ V. (b) The variation of $E_x$ against $t$ for the DM and CM data shown in (a). The phase difference between the two signals is characterised by $\delta t$, as described in the text.  \label{Fig:2}}
\end{figure}

\begin{figure*} 
\includegraphics[angle=0,width=0.16\textwidth]{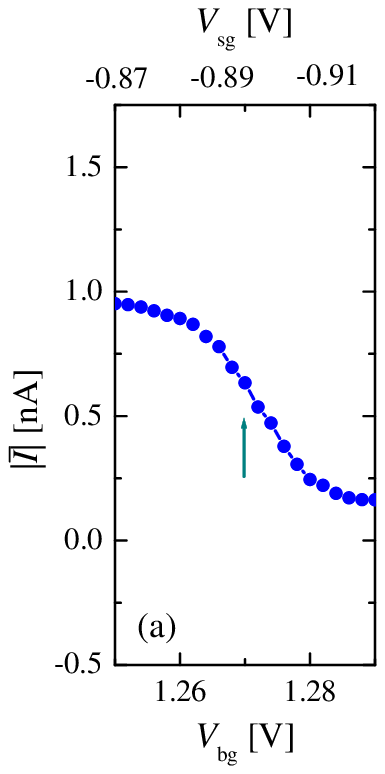}
\includegraphics[angle=0,width=0.32\textwidth]{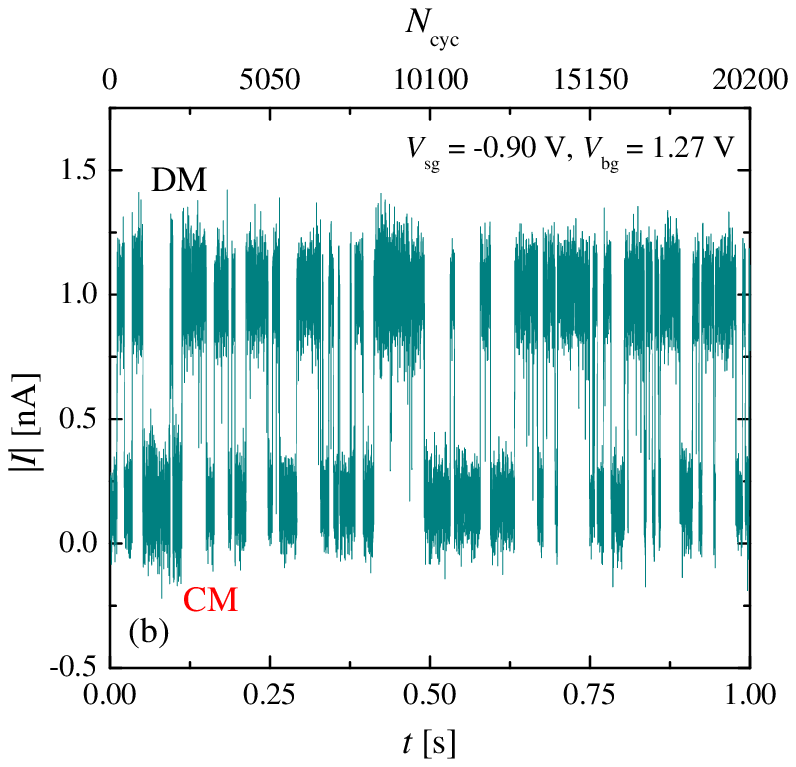}
\includegraphics[angle=0,width=0.16\textwidth]{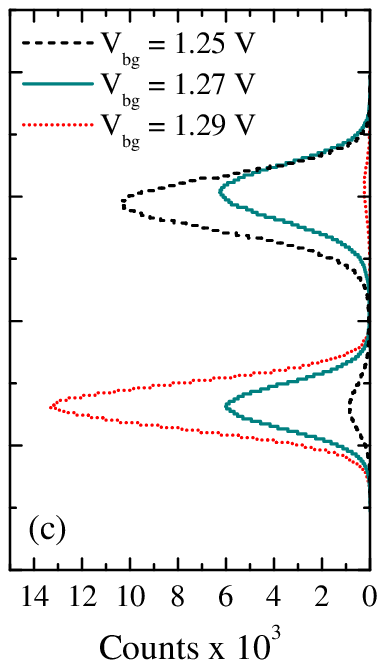}
\includegraphics[angle=0,width=0.32\textwidth]{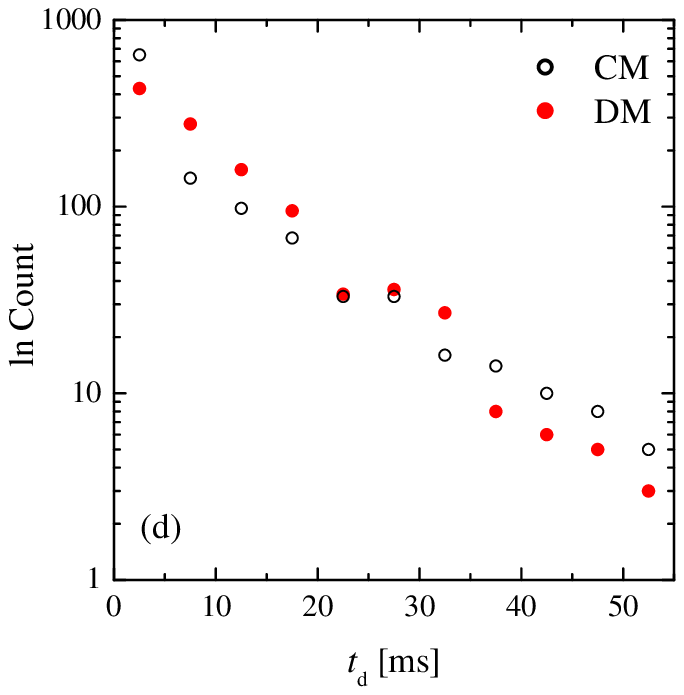}
\caption{(a) $|\bar I|$ against $V_{bg}$ and $V_{sg}$ close to the transition between the DM and CM regimes. The arrow indicates the point at which the time spent in the DM and CM is approximately equal. (b) $|I|$ against $t$ for $V_{sg}=-0.9$ V and $V_{bg}=1.27$ V. The top axis indicates the number of ac cycles. (c) Histograms for the data shown in (b) and for the cases in which the system is mostly in the DM or CM regimes, for $V_{bg} =1.25$ and 1.29 V, respectively. The data points are binned in 10 pA intervals. (d) Distribution of the dwell times $t_d$ recorded for each transport mode over a period of 1.8 s. The values of $t_d$ are binned in 5 ms intervals.  \label{Fig:3}}
\end{figure*} 

The distribution of the WS-DL decoupling threshold has an important consequence when the WS is driven continuously by a sinusoidal voltage, namely the appearance of a bistability in the surface electron conduction. In Fig. 3(a) we show examples of the surface electron current that flows in response to a sinusoidal driving voltage of frequency 20.2 kHz and amplitude 40 mV (peak-to-peak) for two values of $V_{bg}$. Here $V_{sg}=-0.5$ V. In this measurement we employ a current preamplifier (Stanford Research SR570) with low noise characteristics but a reduced bandwidth (200 kHz). Signal averaging over several thousand cycles is again employed to resolve the (average) current. For the lower value of $V_{bg}$, a current peak is observed during each ac half-cycle as the WS decouples from (and then quickly re-couples to) the DL. For the higher value of $V_{bg}$ the electron density, and the pressing electric field perpendicular to the helium surface, are increased and no decoupling occurs. By integrating the current signal, the number of electrons that pass through the central microchannel, and so the voltage drop due to this current flow, can be calculated. This voltage is subtracted from $V_{lr}$ in order to determine $E_x$ at each value of $t$. This is shown in Fig. 3(b). It is clear that the decoupling of the WS results in a drop in $E_x$.

Crucially, there is a difference in phase between the current signals shown in Fig 3(a). This is because, in the case where the WS-DL decoupling occurs, the rapid transfer of electrons from one reservoir to the other (for example from left to right) allows the system to approach electrostatic equilibrium. Then, when the polarity of the driving voltage is reversed, the electrons can quickly begin to move in the opposite direction (right to left). In the case where the decoupling does not occur during the initial half-cycle, a large imbalance of charge still exists even when the polarity of the driving voltage begins to change. This charge imbalance continues to drive the electrons from left to right until the driving voltage becomes large enough to reverse the flow. Thus the behaviour of the system during one half-cycle directly influences the behaviour during the next, which influences the behaviour during the subsequent cycle, and so on. A phase difference is therefore established between the two current waveforms and the system can become `locked' in either of these `decoupling' or `coupled' transport modes (DM and CM, respectively). In the following discussion, we characterise the phase difference between the DM and CM signals by measuring the difference in the time $\delta t$ at which each signal reaches the condition $E_x=0$, as indicated in Fig. 3(b). 

As  $V_{bg}$ increases, the increasing strength of coupling between the electron system and the helium results in a transition from the DM to the CM. However, when the value of $V_{bg}$ is set close to the voltage at which the transition occurs, a spontaneous switching between the two transport modes is observed, at frequencies much lower than that of the ac drive. To study this effect we employed a lock-in amplifier to measure the first harmonic component of the current signal. Although the current response is not sinusoidal in either the DM or CM case, this measurement allows us to record large and small current signals, demodulated from the ac carrier signal, that are related to the root-mean-square amplitudes of the DM and CM current, respectively. The amplitude of this signal, denoted $|I|$, was recorded using a spectrum analyser at an acquisition rate of $f_{aq}=16$ kHz. We denote the time-averaged value of this signal as $|\bar I|$. Working at the carrier frequency of 20.2 kHz offers several advantages. Because our circuit (Fig. 1(a)) consists of a resistor and two capacitors in series, it serves as a high-pass filter. The lock-in technique essentially modulates the low frequency signal due to the DM-CM switching to centre at the carrier frequency, which is high enough to pass through the circuit, and then demodulates to recover the original transition dynamics. Also, modulation to higher frequencies allows the minimisation of the input noise of the current preamplifier \cite{Yehe1700135}.  

In Fig. 4(a) we show the variation of $|\bar I|$ (averaged over 10 seconds) as $V_{bg}$ is swept positive and $V_{sg}$ is swept negative. By changing $V_{bg}$ and $V_{sg}$ simultaneously between the values shown on the upper and lower axes, the number of electron rows formed in the CM remains approximately constant, according to a previous mapping of the structural diagram of the quasi-1D electron system \cite{PhaseDiagram}. However, the electron density and the pressing electric field acting on the electron system change with the electrode bias values. As $V_{bg}$ becomes more positive and $V_{sg}$ becomes more negative, $n_s$ and $E_z$ increase and $|\bar I|$ changes smoothly from a high value to a lower value. 

This smooth change is a result of the telegraph-like switching between the two conductance modes; as the WS-DL coupling increases in strength the system spends less time in the decoupling transport mode and more time in the coupled transport mode, and the average current signal smoothly decreases from a high value to a low value. At the centre of the step in $|\bar I|$ the system spends an equal amount of time in the DM and CM. The time dependence of $|I|$ for this bias condition is shown in Fig. 4(b). The spontaneous switching between the DM and CM is observed. The time spent between switching events is typically much longer than the ac period (49.5 $\mu$s). Histograms of the current measurements are shown in Fig. 4(c). These are determined by binning the current data points in intervals of 10 pA. (Here the Gaussian distribution of $|I|$ values around the large and small values corresponding to the DM and CM results from the white noise of the current preamplifier. This is the limiting noise source in the measurement, neglecting the low-frequency switching between the DM and CM.) For $V_{bg}=1.27$ V the distributions are of almost equal amplitude confirming that the time spent in the DM or CM is approximately equal. By increasing (decreasing) $V_{bg}$ by 20 mV, the WS-DL coupling is strengthened (weakened) and the time spent in the CM (DM) is greatly increased.

Further analysis of this bistability is shown in Fig. 4(d). By monitoring the switching events from the DM to CM, and vice versa, dwell times $t_d$ for both modes were determined. The resolution of the measurement of $t_d$ is $1/f_{aq}$, which is similar to the ac period. In Fig. 4(d) we plot the number of counts falling within 5 ms intervals of $t_d$. The probability of mode switching follows the general exponential form $P(t_d)=P(0)e^{-t_d/\tau_0}$, where $1/\tau_0$ is a constant decay rate, indicating that the probability of mode switching is the same for each ac cycle \cite{Grasser201239,Yehe1700135}. We typically obtain values of $\tau_0\approx1\sim1000$ ms, for the range of temperature and gate electrode bias used here. Although our system is more complex than a single-particle picture, it is instructive to compare the dynamics measured here to a thermally activated process for which $1/\tau_0=f_0e^{-V_B/k_BT}$, where $f_0$ is the attempt frequency and $V_B$ is the effective potential barrier height separating the DM and CM states. Using the value $f_0=10$ GHz, which we expect to be similar to the frequencies associated with the electron motion within its dimple and about its lattice site in the WS \cite{MonarkhaandKono,Peeters1DCrystal}, we estimate $V_B\approx20$ K. 

\section{V. Discussion}

The rate at which electrons can pass through the central microchannel depends strongly on a variety of factors that change as the values of both $V_{bg}$ or $V_{sg}$ are varied \cite{PhaseDiagram}. The electron density, pressing electric field, effective channel width, strength of the lateral electrostatic confinement and the commensurability of the electron lattice within the quasi-1D confinement are all dependent on the electrode bias values \cite{PhysRevB.82.201104,PhysRevLett.109.236802,*Ikegami2015}. Each of these parameters can influence the WS-DL decoupling threshold force. The conductance of the electron system when the WS is both coupled to and decoupled from the DL, and therefore the rate at which electrons can move between the reservoirs, is also influenced by these factors. We therefore expect the WS-DL decoupling threshold to be strongly influenced by the sample bias conditions. This influence is, generally, difficult to understand quantitatively and a full analysis of the ac response of the WS in our device with changing bias conditions is beyond the scope of this report. However, it is instructive to measure the value of $\delta t$ under different bias conditions and compare it with the rate of switching between the DM and CM. In Fig. 5(b) we show the average time spent in each conductance mode $\bar{t}_d$, and the value of $\delta t$, as recorded at $V_{bg}$ and $V_{sg}$ bias values close to the DM-CM transition. The points in the $V_{sg}$-$V_{bg}$ plane at which the DM and CM waveforms were recorded in order to determine $\delta t$ are shown in Fig. 5(a). For each value of $V_{sg}$, the value of $V_{bg}$ was adjusted either side of the bistable region such that no switching behaviour was observed, ensuring that the system lay fully in either the DM or CM regime. The value of $\delta t$ was determined using the waveforms recorded at these bias points. Values of $\bar{t}_d$ were recorded at values of $V_{sg}$ and $V_{bg}$ at the boundary between the DM and CM regimes, for which the time spent in either state was approximately equal.

\begin{figure} 
\includegraphics[angle=0,width=0.48\textwidth]{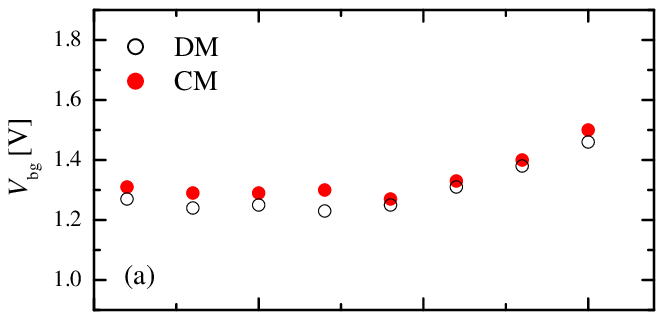}
\includegraphics[angle=0,width=0.48\textwidth]{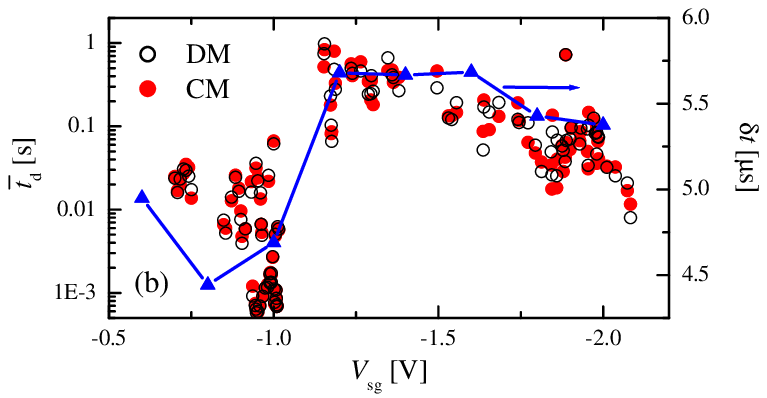}
\caption{(a) Values of $V_{bg}$ at which the DM or CM waveforms were measured against $V_{sg}$. (b) The average dwell time $\bar{t}_d$ and $\delta t$ against $V_{sg}$. The values of $\bar{t}_d$ for the DM and CM are approximately equal indicating that the system spends an approximately equal time in each conductance mode.  \label{Fig:3}}
\end{figure} 

There is a clear correspondence between $\bar{t}_d$ and $\delta t$; when $\delta t$ is small the system readily switches between the DM and CM, resulting in small $\bar{t}_d$. When $\delta t$ is large the system remains in either conductance mode for longer periods. The dependence of $\bar{t}_d$ on $\delta t$ is exponentially strong; the typical dwell time varies over 4 orders of magnitude as $\delta t$ varies by up to a factor 2. We attribute this effect to the Gaussian-like distribution of the WS-DL decoupling threshold force. For example, for the DM case, a decoupling event for which the threshold force is much higher than the average value results in a delay in the decoupling. If this delay is long enough then the decoupling does not occur at all during the ac half-cycle, and the system switches to the CM. Similarly, a switch from the CM to the DM can occur due to a decoupling event that occurs with a threshold force much lower than the average value. The switching events become rare when threshold force values occupying the far tails of the Gaussian distribution are required to induce the transition. This results in the low-frequency bistable behaviour. This is confirmed by the quantitative analysis shown here of both the Gaussian distribution of the decoupling threshold force and the delay between the DM and CM waveforms; the typical values of $\delta t$ are several times greater than $\sigma_t$ and only a small fraction of the decoupling events can induce DM-CM switching. At the Gaussian tails, the quasi-exponential dependence of the number of decoupling events with changing force results in the strong dependence of $\bar{t}_d$ on $\delta t$.

The observation of a distribution in the WS-DL decoupling threshold force is not surprising for the quasi-1D case examined here. Many experimental studies and simulations have demonstrated that geometrical confinement results in the formation of irregular quasi-1D lattice arrangements containing structural defects \cite{Peeters1DCrystal,PhysRevB.84.024117,PackingAndMelting,PhysRevLett.109.236802,*Ikegami2015}. The irregular electron arrangement can impede the constructive interference that deepens the DL during BC scattering. In addition, because the defected lattice states can have many possible configurations separated by small energies, fluctuations between different lattice states occur spontaneously at finite temperature, which can also influence the DL formation. Furthermore, in our experiment the central microchannel is continuously repopulated with electrons as the current flows, and we expect the electron lattice to exhibit different configurations as it enters and moves along the microchannel. Because the DL depth depends sensitively on these factors, and the decoupling threshold force depends on the DL depth, a distribution in the decoupling threshold is not unexpected. 

External factors may, in principle, also influence the WS-DL decoupling; mechanical vibrations may lead to oscillations in the helium surface height, and so the pressing electric field exerted on the electron system, which would influence the WS-DL decoupling. More directly, the excitation of surface waves might produce an interference which reduces the depth of the DL. Electrical noise from the sample electrodes could also influence the WS-DL decoupling. However, we have found no evidence of these external factors influencing the WS-DL decoupling or the DM-CM bistability. The dilution refrigerator used in the measurement is cooled by a pulse-tube cooler, which should be the largest source of mechanical noise in our experimental set-up. On switching off the pulse-tube cooler for short periods, during which $T$ remains constant, no change in the decoupling behaviour shown in Fig. 2 or in the values of $\bar{t}_d$ was recorded. Attempts to improve the filtering of electrical noise entering the sample cell, or to increase the noise level by adding large resistances to the dc bias lines, also resulted in no changes to the observed behaviour. The spectrum of the mechanical and electrical noise exhibit maxima at certain characteristic frequencies, for example at 60 Hz in the case the electrical noise spectrum; however, no excess switching was observed at these frequencies. We conclude that the distribution in the WS-DL decoupling threshold arises due to intrinsic structural disorder in the confined electron system and not from any external sources.

\section{VI. Temperature and Commensurability Dependence of the Transport Mode Switching}

The electron motion that influences the WS-DL coupling is thermally driven, and it is therefore of obvious interest to investigate the dependence of the transport bistability on temperature. Also, it is well-known that for quasi-1D WS systems the commensurability of the electron lattice with the confining potential influences the degree of electron positional order and fluctuations in electron lattice configurations\cite{Peeters1DCrystal}. When the lattice is commensurate with the confinement the system is highly ordered into rows. When the lattice is incommensurate with the confinement, structural defects appear in order to reduce the total energy of the electron system and fluctuations between metastable lattice configurations of similar energy become more common. In this case, an increase in the DM-CM switching rate should be expected. 

The bistable behaviour observed here can be investigated at different temperatures, and for varying degrees of lattice commensurability. However, because the average decoupling threshold depends strongly on these parameters, their variation requires that the electrode bias conditions be changed in order to to remain in the bistable transport regime. As described above, the modification of the bias conditions can result in changes in many factors which influence the rate at which electrons can move between the reservoirs, and so the value of $\delta t$. For the thermally activated process mentioned above, this is equivalent to a change in the value of $V_B$. In this section we demonstrate that, due to this effect, it is not straightforward to unambiguously determine the influence of temperature and commensurability on the decoupling threshold distribution.

\begin{figure} 
\includegraphics[angle=0,width=0.45\textwidth]{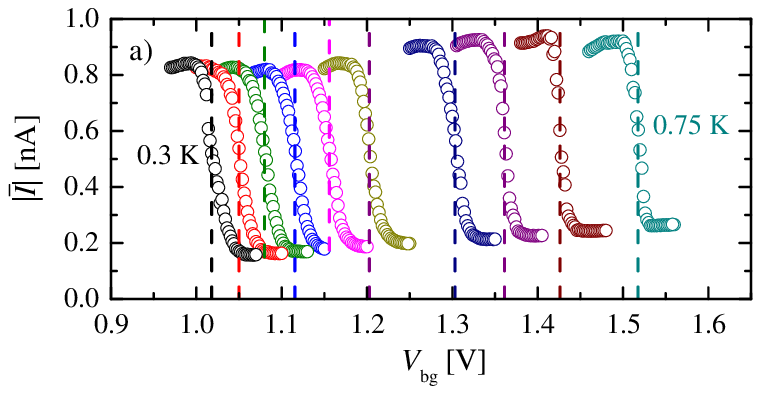}
\includegraphics[angle=0,width=0.45\textwidth]{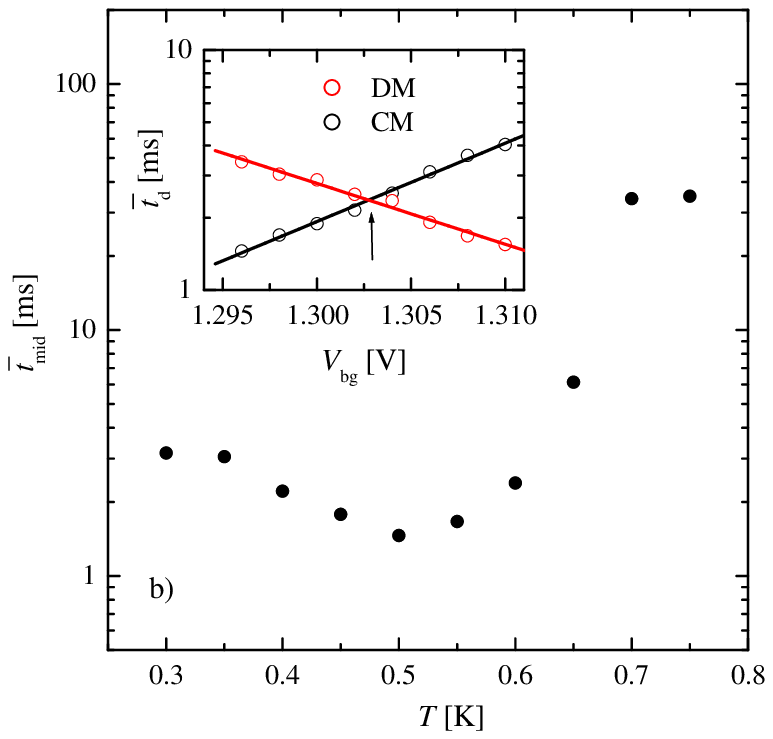}
\caption{(a) $|\bar I|$ against $V_{bg}$ for temperatures ranging from 0.3 K to 0.75 K in 50 mK steps. The dashed lines indicate the values of $V_{bg}$ at which $\bar t_{mid}$ was measured. (b) $\bar t_{mid}$ against T. Inset:  $\bar{t}_d$ for the DM and CM against $V_{bg}$. The solid lines are polynomial fits to the data points. $\bar t_{mid}$ is defined as the value for which the polynomial functions are equal, as indicated by the arrow. \label{Fig:1}}
\end{figure} 

In Fig. 6(a) we show $|\bar I|$ against $V_{bg}$ at different temperatures, for $V_{sg}=-0.6$ V, in the bistable transport regime. The threshold between the DM and CM regimes depends strongly on the temperature, increasing from $V_{bg}=1.02$ V at $T=0.3$ K to $V_{bg}=1.52$ V at $T=0.75$ K. This is due to the thermal motion of electrons about their lattice sites. This motion increases with temperature, degrading the structure of the dimple lattice; a higher value of $V_{bg}$ is therefore required to maintain the coupling of the WS to the DL as the temperature increases. It is also clear that the magnitude of the current step changes with temperature as the average conductance, in both the CM and DM modes, changes with the temperature and bias conditions. The current step is smallest for $T\approx0.5$ K, and is larger at the lowest and highest temperatures.  

In Fig. 6(b) we show the average dwell time at the centre of the current steps against $T$. Here we find the average dwell time by measuring $\bar t_d$ for both the CM and DM as $V_{bg}$ is varied and, by extrapolation of the resulting data points using a polynomial line fitting, find the value for which the time spent in the decoupling and coupling modes is exactly equal. An example is shown in the inset to Fig. 6(b). As expected, the value of $V_{bg}$ at which this condition is satisfied is at the centre of the current step, as indicated by the dashed lines in Fig. 6(a). We denote the average dwell time found by this method as $\bar t_{mid}$. It is clear that the value of $\bar t_{mid}$ exhibits a correspondence with the size of the current step. When the step size is large the dwell time is also large. This correspondence can be attributed to the dependence of $\delta t$ on the difference in the relative magnitudes of the (average) conductance in the decoupling and coupled transport modes. When the difference is large, the phase difference between the current signals of the CM and DM is also large, making switching events rarer. It is not straightforward to change $T$ and also keep $\delta t$ constant. We conclude, therefore, that the changes in the bias conditions necessary to observe the DM-CM bistability as the temperature changes make it difficult to gain information about the influence of temperature on the distribution of the decoupling threshold force. 

The dependence of the current bistability on the structural order of the quasi-1D lattice can be investigated by changing the gate electrode voltages in small steps in order to smoothly vary the electron density in the central microchannel. As $n_s$ changes, so does the average number of electron rows $N_y$ forming the quasi-1D lattice \cite{PhysRevLett.109.236802,*Ikegami2015,Stick-slip}. When the number of lattice spacings across the electron crystal is commensurate with the effective channel width, the electron system is highly ordered, forming a discrete number of rows ($N_y$ is integer). For incommensurate cases the electron lattice becomes strongly disordered (in this case the value of $N_y$, calculated using our finite element model, is non-integer) \cite{beysengulov2016structural}.

\begin{figure} 
\includegraphics[angle=0,width=0.4\textwidth]{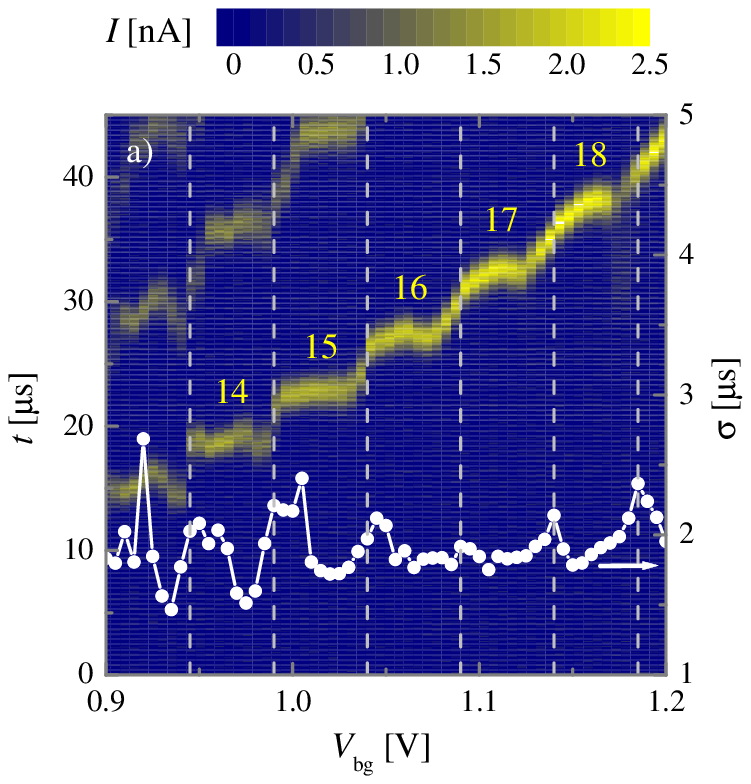}
\includegraphics[angle=0,width=0.4\textwidth]{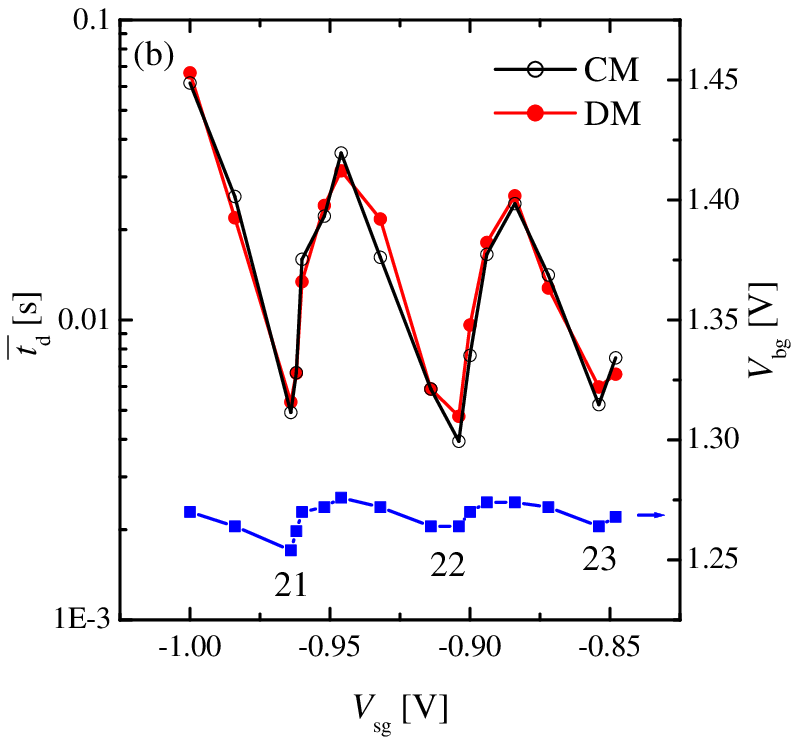}
\caption{(a) $I$ against $t$ and $V_{bg}$, and the standard deviation $\sigma$ found by fitting a Gaussian function to the current peaks, against $V_{bg}$. Here $V_{sg}=-1$ V. Values of $N_y$ obtained from FEM analysis are shown on the plot. The dashed lines are a guide to the eye, marking the points at which structural transitions between successive electron row configurations occur ($N_y$ is half-integer). (b) $\bar t_d$ against $V_{sg}$ for the DM and CM. The values are measured at points close to the centre of the bistable regime, which is achieved by varying $V_{bg}$ at each value of $V_{sg}$. The values of $V_{bg}$ are shown by the square symbols. $N_y$ is close to integer at points where $V_{bg}$ is smallest, as shown; here $N_y$ varies from 21 to 23. \label{Fig:1}}
\end{figure} 

In Fig. 7(a) we show $I$ recorded during $V_{lr}$ ramps and for different values of $V_{bg}$. These measurements are similar to those shown in Fig. 2(a). Here $V_{sg}=-1$ V. The current peaks due to the WS-DL decoupling generally occur at later times as $V_{bg}$, and so the strength of the WS-DL coupling, increases. However, the time at which the decoupling occurs ($t_p$) does not increase monotonically but in a series of steps, due to the varying positional electron order as $N_y$ changes \cite{Stick-slip}. At the centre of each plateau-like feature $N_y$ is close to integer, whilst each step-like change in $t_p$ corresponds to a structural transition. $N_y$ values calculated using FEM are shown on the plot; here $N_y$ varies from some 14 to 18 rows. To investigate the influence of the lattice defects that are expected to proliferate at each structural transition on the WS-DL decoupling threshold, the width of each current peak was determined by fitting a Gaussian function with standard deviation $\sigma$ to the current peak at each value of $V_{bg}$. Because the current peaks are very close to Gaussian (as shown in Fig. 2), this direct fitting procedure is a more straightforward way in which to characterise the width of the decoupling threshold force distribution than by using a model current peak in order to calculate $\sigma_t$, as was performed above. The values of $\sigma$ are shown in Fig. 7(a). A dependence of $\sigma$ on the positional electron order is observed; an increase in $\sigma$ close to each structural transition indicates that the distribution of the WS-DL decoupling threshold increases in width, as expected.

The increase in the width of the decoupling threshold distribution close to each structural transition should result in an increased probability of switching between the DM and CM modes, and so a decrease in the dwell times $\bar{t}_d$. However, this is difficult to determine experimentally because the average DM-CM decoupling threshold oscillates in the $V_{bg}$-$V_{sg}$ plane as $N_y$ changes. This is shown in Fig. 7(b) (square symbols), and arises due to the weakened WS-DL coupling due to positional disorder for the incommensurate lattice states. As described above, the electron transport properties are influenced by the varying bias conditions. In Fig. 7(b) we also show the variation of $\bar{t}_d$ as the bias conditions, and $N_y$, are changed. The value of $\bar{t}_d$ is smaller for the commensurate states. This is contrary to the expectation that the DM-CM switching should become less frequent in this case, due to the increased order of the electron lattice. We attribute this to the influence of the WS-DL coupling strength on $\delta t$. In the DM, for commensurate states in which the average decoupling threshold force becomes larger, the decoupling occurs later in the ac cycle. This will result in a decrease in $\delta t$, and so a decrease in $\bar{t}_d$. We conclude that it is difficult to unambiguously study the influence of the lattice commensurability on the width of the decoupling threshold distribution by monitoring the DM-CM switching rate. We note that the strong influence of the lattice commensurability on $\bar{t}_d$ can also explain the scatter in the points shown in Fig. 5(b). These values were measured over a wide range of $V_{bg}$ and $V_{sg}$ values, with a large variation in the number of electron rows and, for each data point, the degree of commensurability.

\section{VII. Conclusions}

We have demonstrated that the decoupling of a quasi-1D electron crystal from the dimple lattice formed at the surface of liquid helium exhibits a Gaussian-like distribution in threshold force. By applying a continuous sinusoidal driving voltage to the system, a bistable regime is observed in which the system spontaneously switches between conductance modes in which decoupling does or does not occur during each ac cycle. In this regime, the switching rate between the DM and CM is extremely sensitive to the WS transport properties. We have made a quantitative analysis of the transport characteristics of the electron system, which demonstrates that the low-frequency current fluctuations are driven by decoupling events with threshold values that occupy the far tails of the Gaussian distribution. We have also shown that the low-frequency switching is exponentially sensitive to the WS transport properties. Our experiment provides an interesting example of a system in which high-frequency electron dynamics can give rise to low-frequency current fluctuations that appear in response to continuous driving. The measurement techniques demonstrated here could be useful in studying nonlinear carrier dynamics in other strongly-correlated electron systems.     

\begin{acknowledgments}
We thank N. R. Beysengulov for finite element modelling of the microchannel device. This work was supported by the Taiwan Ministry of Science and Technology (MOST) through Grants No. MOST 103-2112-M-009-001 and No. MOST 104-2112-M-009-022- MY3 and the Taiwan Ministry of Education (MOE) ATU Plan, and by JSPS KAKENHI Grant No. 24000007 and JP17H01145. This work was performed according to the Russian Government Program of Competitive Growth of Kazan Federal University.

\end{acknowledgments}

D. G. R. and S.-S. Y. contributed equally to this work.


%

\end{document}